# Comparison of retinal regions-of-interest imaged by OCT for the classification of intermediate AMD

Danilo A. Jesus[1,2], Eric F. Thee[2,3], Tim Doekemeijer[1], Daniel Luttikhuizen[2,3], Caroline Klaver[2,3,4,5], Stefan Klein[1], Theo van Walsum[1], Hans Vingerling[2], and Luisa Sanchez[1,2]

[1]Biomedical Imaging Group Rotterdam, Department of Radiology & Nuclear Medicine, Erasmus MC, Rotterdam, The Netherlands;
[2]Department of Ophthalmology, Erasmus MC, Rotterdam, The Netherlands;
[3]Department of Epidemiology, Erasmus MC, Rotterdam, The Netherlands;
[4]Department of Ophthalmology, Radboud University Medical Center, Nijmegen, The Netherlands;
[5]Institute of Molecular and Clinical Ophthalmology, Basel, Switzerland.

# ABSTRACT

**Purpose:** To study whether it is possible to differentiate intermediate age-related macular degeneration (AMD) from healthy controls using certain regions of interest (ROIs) in optical coherence tomography (OCT) data, in particular the choroid.

**Methods:** A total of 15744 B-scans (Bioptigen SD-OCT, NC, USA) from 269 intermediate AMD patients and 115 normal subjects, selected from a publicly available dataset, were used in this study (split on subject level in 80% train, 10% validation and 10% test). From each OCT B-scan, three ROIs were extracted: the entire retina, the retinal pigment epithelium (RPE)/Bruch's membrane (BM) complex, and the choroid (CHO). These ROIs were obtained using two different methods: masking and cropping. In addition to the ROIs, the whole OCT B-scan and the binary mask corresponding to the segmentation of the RPE-BM complex were used. For each subset, a convolutional neural network (based on VGG16 architecture and pre-trained on ImageNet) was trained and tested. The performance of the models was evaluated using the area under the receiver operating characteristic (AUROC), accuracy, sensitivity, and specificity.

**Results:** All trained models presented an AUROC, accuracy, sensitivity, and specificity equal to or higher than 0.884, 0.816, 0.685, and 0.644, respectively. The model trained on the whole OCT B-scan presented the best performance (AUROC = 0.983, accuracy = 0.927, sensitivity = 0.862, specificity = 0.913). The models trained on the ROIs obtained with the cropping method led to significantly higher outcomes than those obtained with masking, except for the retinal tissue, where no statistically significant difference was observed between cropping and masking (p = 0.47). The binary mask of the RPE-BM complex was enough to obtain an AUROC of 0.932.

**Conclusion:** While using the complete OCT B-scan provided the highest accuracy in classifying intermediate AMD, models trained on limited information such as the choroid can still achieve high performance.

# 1 Introduction

Age-related macular degeneration (AMD) is the leading cause of blindness in the elderly of European descent (Colijn et al. 2017). The disease primarily affects the central part of the retina, the macula, which provides sharp visual perception. Early and intermediate stages of the disease generally develop without any symptoms and are characterized by drusen (lipoprotein deposits) and pigment changes in the retina. The disease can further progress towards an end-stage characterized by geographic atrophy (GA, dry AMD) and/or choroidal neovascularization (CNV, wet AMD), both often causing severe visual impairment. So far, treatment is only available for patients with CNV. Recent reports have shown that nutritional supplements and changes in lifestyle and diet can reduce the risk of progression towards late stages of AMD by at least 50% (de Koning-Backus et al. 2019, Merle et al. 2019, Chew et al. 2014). Monitoring the disease progression is therefore important.

AMD is diagnosed and monitored by the combined evaluation of colour fundus photography (CFP) and optical coherence tomography (OCT) scans. Nevertheless, most AMD severity classifications are based on CFP evaluation only (Thee et al. 2020, Saha et al. 2019, Hussain et al. 2018). Evaluation of OCT B-scans is more time-consuming and error-prone, as it requires navigation through many images. Information on OCT scans, however, can easily be extracted by automatic image processing approaches, such as applying deep learning (DL) techniques.

Lee et al. (Lee et al. 2017) trained a modified version of the VGG16 (Simonyan & Zisserman 2014) convolutional neural network (CNN) to categorize OCT images as either normal or AMD. At image level, they achieved an area under the receiver operating characteristic (AUROC) of 92.78% with an accuracy of 87.63%. Similar results were obtained by Yoo et al. (Yoo et al. 2019) who, using a pre-trained VGG19 (Simonyan & Zisserman 2014) on OCT alone, showed diagnostic effectivity with AUROC of 90.6% and 82.6% accuracy. More recently, Russakoff et al. (Russakoff et al. 2019) presented a new architecture, AMDnet, for predicting the likelihood of converting from early/intermediate to advanced wet AMD, achieving an AUROC of 89%.

Preprocessing techniques have been shown to be useful in increasing the performance of DL models. In the work presented by Russakoff et al. (Russakoff et al. 2019), each B-scan was cropped from the ILM to a fixed offset below Bruch's membrane (BM), which included the choroidal information over a fixed area beneath the choriocapillaris. Using this restricted region of interest (ROI), the authors reported an improvement on their baseline approach, a VGG16 model, from 66% to 82% AUROC. Whereas this preprocessing was performed to reduce the variance of the training set and create some invariance to scale, it is unclear whether the accuracy of the classification could be improved by, for instance, selecting different ROIs within the image. Recently, Srivastava et al. (Srivastava et al. 2020), after comparing intermediate-stage AMD classification in OCT imaging with and without the choroid, observed a higher AUROC in the model trained with the retina and choroidal information. However, no statistical analysis was performed in their study, which prevents further conclusions on the added value of the choroid for the intermediate AMD classification.

Although early and intermediate stages of AMD have been associated with the presence of soft drusen within the retinal pigment epithelium (RPE) and BM complex, studies have shown that variable choroidal thickness exists among patients with the clinical diagnosis of wet and dry AMD (Manjunath et al. 2011). Recently, a systematic review (Farazdaghi & Ebrahimi 2019) suggested that choroid undergoes changes in the earlier stages of AMD. However, some aspects remain unclear, such as the significance of choroidal information with respect to visual function and disease progression over time, and whether the choroidal data retrieved from an OCT image would be enough to differentiate between healthy controls and intermediate AMD. Further research into the role of choroidal information from OCT imaging in AMD, including its potential as a biomarker for disease progression, could lead to improved diagnostic and treatment options for individuals with the disease.

Therefore, in this work, we investigate whether DL models trained on a limited region of the OCT data can accurately classify intermediate AMD from healthy controls. Besides looking at conventional clinically relevant ROIs, we analysed the information in the choroidal region and investigated whether a DL model can identify patterns in the data that are not visually or clinically evident.

## 2 Methods

The pipeline for this work is depicted in Figure 1. The computations were performed at B-scan level in the dataset described in Section 2.1. To analyse the ROIs, the images were masked following the approaches described in Section 2.2. Next, a CNN was trained per ROI, as described in Section 2.3. Finally, the models were evaluated, as described in Section 2.4.

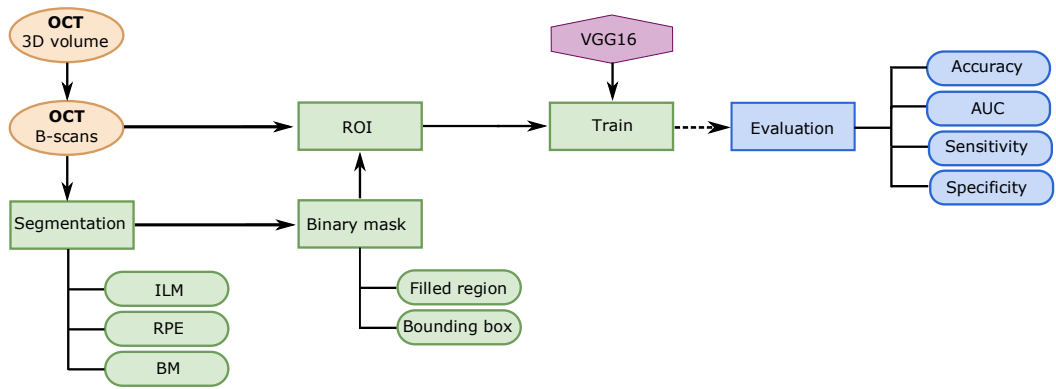

**Figure 1.** Flowchart of the steps taken in this work. ILM: Inner Limiting Membrane; RPE: Retinal Pigmented Epithelium; BM: Bruch's Membrane.

### 2.1 Dataset

For this study, a publicly available OCT dataset (Farsiu et al. 2014) acquired with SD-OCT imaging systems from Bioptigen, Inc (Research Triangle Park, NC) was used. The images consisted of eyes of subjects between 50 and 85 years of age, who clearly exhibited intermediate AMD with large drusen (>125 $\mu m$). The control subjects had no evidence of macular drusen or AMD in either eye at the baseline visit or in the follow-up years. Rectangular volumes centred at the fovea were captured with 1000 A-scans per B-scan and 100 B-scans per volume, which resulted in 38400 OCT B-scans from 269 AMD patients and 115 healthy subjects. Although each OCT volume consisted of 100 B-scans, it has been determined that only the B-scans in the middle were relevant for the classification task, as these images are located at the macula, the most affected region in AMD. This resulted in the used data consisting of a total of 15572 images (B-scans), 10873 AMD and 4699 healthy, and it was divided as follows: 80% of the data for training (8697 AMD images, 3756 control images), 10% of for validation (1083 AMD images and 451 control images), and 10% for test (1093 AMD images and 492 control images). The division was done at subject level to avoid bias, which justifies the small discrepancies in number between validation and test sets.

### 2.2 Regions of interest

In addition to the standard OCT B-scan image (IMG), different ROIs were defined in this study. The selection of the ROIs was assisted by the segmentation of three retinal layers: inner limiting membrane (ILM), retinal pigment epithelium (RPE), and Bruch's membrane (BM), as described in a previous work (De Jesus et al. 2020). With these layers as a reference, three ROIs were defined: retinal tissue between the ILM and the BM (ILM-BM), the complex between the RPE and the BM (RPE-BM), and the region below the BM, including information from the choroid (BM-CHO).

In order to prepare the ROIs for training the DL models, two different approaches were explored: (1) masking the exact ROI, and (2) using a bounding box to crop a rectangular area around the ROI. Masking has the advantage of including position and shape information, while cropping ensures that the entire image has pixel information (i.e., no zero intensity pixels). Also, in both methods, the masked/cropped image was resized to 224×224 pixels to fit the standard input of the VGG16 pre-trained model. Figure 2 shows the three images obtained from a standard OCT B-scan after masking and resizing the ROI.

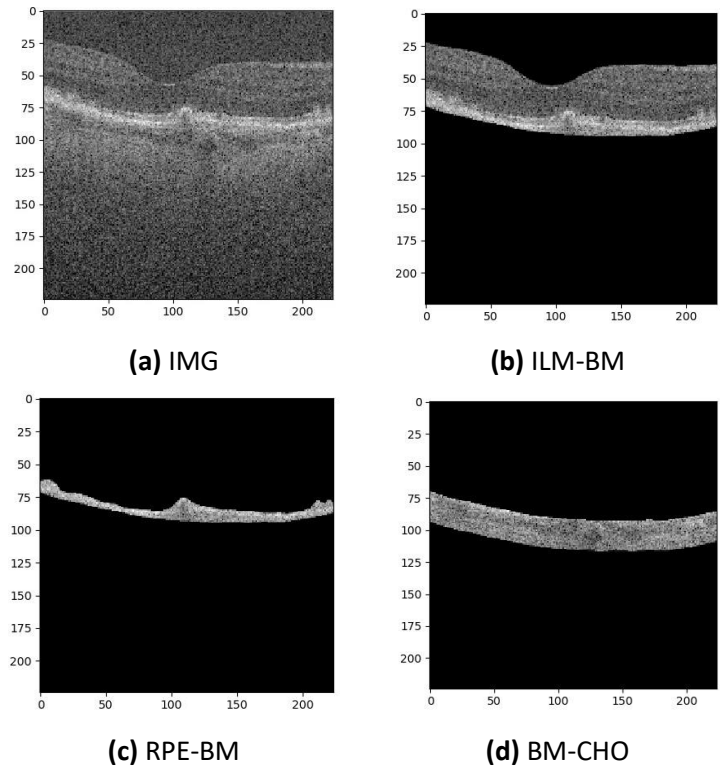

(a) IMG  (b) ILM-BM

(c) RPE-BM  (d) BM-CHO

**Figure 2.** Regions of interest using the masking approach. The numbers refer to the dimension of the images in pixels.

Given the retinal curvature, cropping the standard OCT B-scan using a rectangular shape would lead to images with substantial information from the surroundings of the ROI. This issue was particularly severe in the choroidal region. Hence, retinal curvature correction was done prior to the cropping by flattening the retina. The mean vertical positions of the segmented BM in the image were computed, and then the differences from each BM point to that mean were obtained. These differences were finally used as a reference for the other layers and for the B-scan itself. Every value of the ILM and the RPE was taken and adjusted according to the new flattened BM. The images were altered by shifting every A-scan (column of pixels) according to the distance between the BM and the reference line. Figure 3 shows a standard OCT B-scan (a) and the corresponding image corrected for retinal curvature (b).

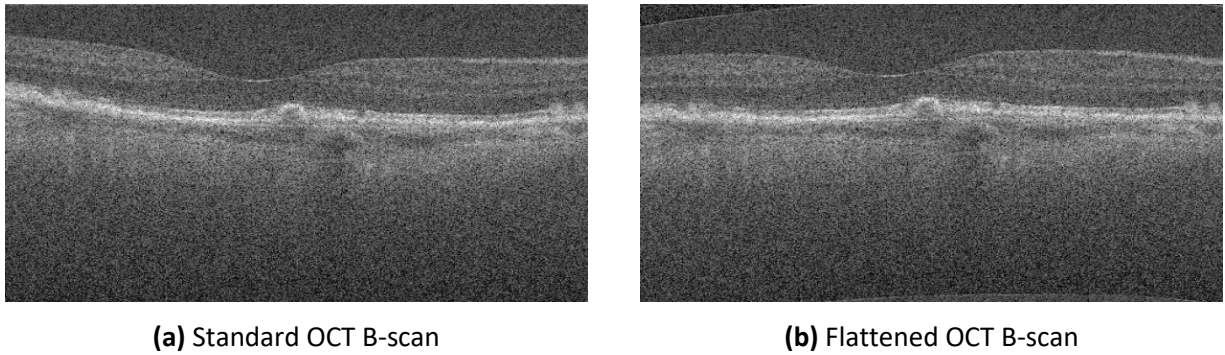

**(a)** Standard OCT B-scan  **(b)** Flattened OCT B-scan

**Figure 3.** OCT B-scan before (a) and after (b) BM flattening.

Figure 4 shows the four ROIs obtained after cropping and resizing the flattened images. The BM-CHO region has been defined from the flattened BM to a fixed offset of 80 pixels into the choroid. This value was estimated based on choroidal thickness in healthy subjects and the axial resolution of the SD-OCT imaging systems from Bioptigen, Inc. (Entezari et al. 2018, Khanifar & Farsiu 2008).

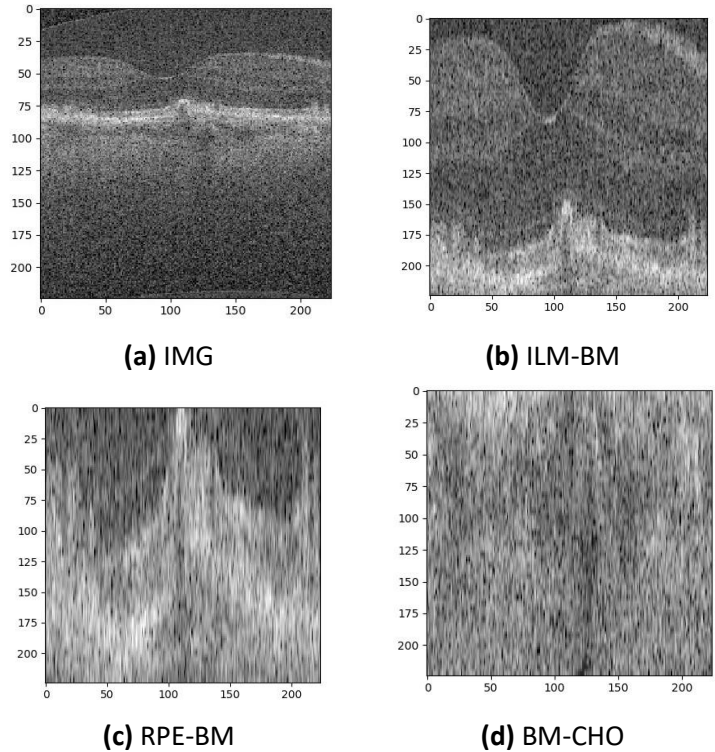

**(a)** IMG  **(b)** ILM-BM

**(c)** RPE-BM  **(d)** BM-CHO

**Figure 4.** Regions of interest using the cropping approach. The numbers refer to the dimension of the images in pixels.

Lastly, the importance of the intensity information in the RPE-BM complex has been weighted by comparing the RPE-BM mask with the masked region. Figure 5 shows an example of the RPE-BM mask for a healthy (a) and intermediate AMD subject (b).

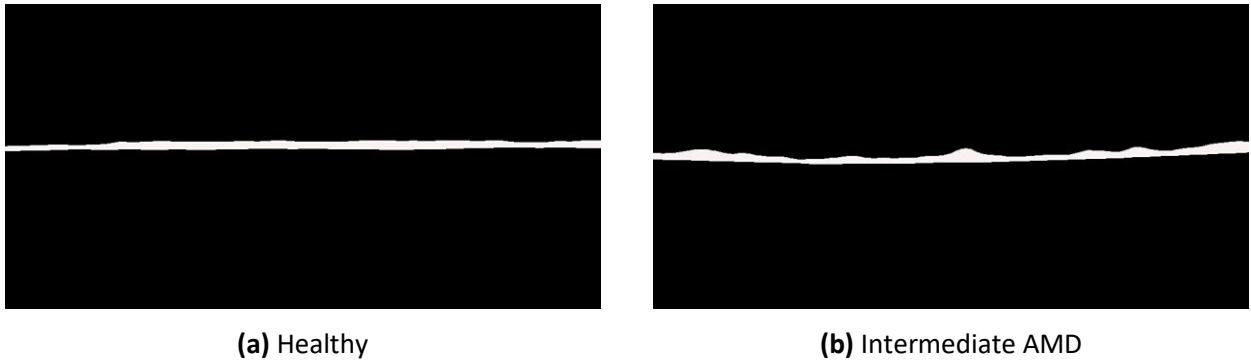

**(a)** Healthy          **(b)** Intermediate AMD

**Figure 5.** RPE-BM mask of a healthy control (a) and intermediate AMD subject (b).

## 2.3 CNN model

One of the most widely used deep CNNs for classification problems is the VGG16 (Simonyan & Zisserman 2014). This architecture can be divided in five blocks, which are formed by a number of convolutional layers followed by a max pooling layer. The first two blocks consist of only two convolutional layers, whereas the following three blocks have three convolutional layers, as can be observed in Figure 6. The final layers of the model are fully-connected layers. A VGG16 model pre-trained on ImageNet (Deng et al. 2009) was fine-tuned, retraining only the fully-connected layers. The only changes to the original architecture were the number of neurons and activation function in the final fully-connected dense layer, which were changed from 1000 to 2, and from softmax to sigmoid, respectively, for binary classification purposes. A batch size of 32 was used.

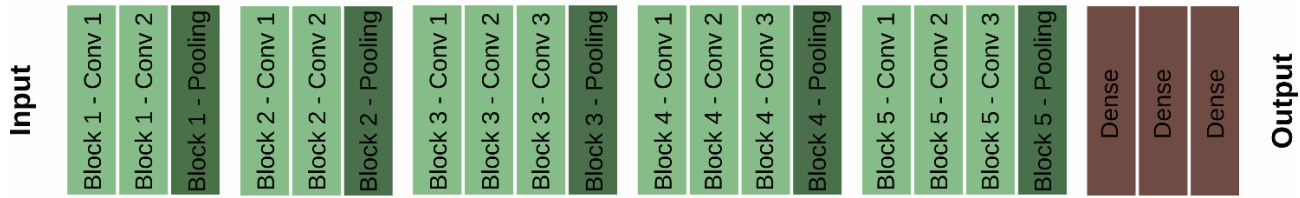

**Figure 6.** VGG16 network architecture. The weights of the convolutional blocks (in green) were frozen, while the dense layers were trained.

In order to make the model more robust, data augmentation was performed, using the following parameters: randomized rotation of the image between 0 and 10 degrees, horizontal flip of the image, brightness adjustments (40% to 120% of the original brightness), shifts in both directions (up to a maximum of 5% of the image pixels on a given direction), and zoom (90% to 120% of the original image size). The values of these augmentations have been chosen to resemble realistic OCT images.

Experiments with different values of hyperparameters were conducted on the model trained on the whole B-scan image to find the optimal values. The stochastic gradient descent with Nesterov momentum and a learning rate of 1e-6 was used as the optimizer, and the binary cross-entropy as the loss function. The training ran for a maximum of 2500 epochs but was stopped as soon as the validation loss did not improve in 20 epochs. These hyperparameters were kept constant for all the other experiments, to compare the effect of having different ROIs in the exact same configuration.

**2.4 Evaluation and interpretability**

The AUROC, accuracy, sensitivity, and specificity of the classification between healthy controls and intermediate AMD subjects were computed for each trained model. The 95 % confidence interval of each metric was obtained with stratified bootstrap resampling (DiCiccio & Efron 1996). The comparison between the ROC curves was performed with the DeLong test (DeLong et al. 1988, Sun & Xu 2014). For all tests, $p < 0.05$ was used to determine statistical significance.

**3 Results**

The accuracy and the loss for the training and validation on the whole B-scan are shown in Figure 7. Table 1 summarizes the numeric results of AUROC, accuracy, sensitivity, and specificity for the test set in all the models trained in the different ROIs (the whole OCT B-scans, the areas segmented according to the masking and cropping approaches, and the RPE-BM mask).

**Table 1.** Performance metrics of all trained models in the test set, ranked by colour. The metrics include AUROC, accuracy, sensitivity, and specificity. The highest value for each metric is highlighted in bold.

|  |  | AUROC | Accuracy | Sensitivity | Specificity |
|---|---|---|---|---|---|
| Whole Image | IMG | **0.983 [0.979 - 0.987]** | **0.927 [0.917 – 0.938]** | 0.862 [0.839 – 0.885] | **0.913 [0.891 – 0.932]** |
| Masking | ILM-BM | 0.970 [0.964 - 0.976] | 0.925 [0.915 - 0.936] | **0.864 [0.839 - 0.888]** | 0.900 [0.878 - 0.920] |
|  | RPE-BM | 0.943 [0.933 - 0.951] | 0.868 [0.855 - 0.882] | 0.848 [0.818 - 0.875] | 0.701 [0.670 - 0.734] |
|  | BM-CHO | 0.884 [0.870 - 0.899] | 0.816 [0.801 - 0.833] | 0.732 [0.697 - 0.767] | 0.644 [0.608 - 0.681] |
| Cropping | ILM-BM | 0.974 [0.968 - 0.979] | 0.917 [0.905 - 0.929] | 0.861 [0.836 - 0.887] | 0.872 [0.846 - 0.896] |
|  | RPE-BM | 0.958 [0.950 - 0.965] | 0.891 [0.878 - 0.905] | 0.862 [0.835 - 0.888] | 0.772 [0.739 - 0.805] |
|  | BM-CHO | 0.915 [0.902 - 0.928] | 0.832 [0.817 - 0.847] | 0.685 [0.653 - 0.716] | 0.852 [0.826 - 0.878] |
| Segmentation only | RPE-BM | 0.932 [0.921 - 0.941] | 0.855 [0.840 - 0.870] | 0.803 [0.772 - 0.834] | 0.705 [0.674 - 0.741] |

The respective AUROC curves are shown in Figure 8. The binary classification for the whole B-scan obtained the highest AUROC (0.983), accuracy (0.927), and specificity (0.913) in classifying intermediate AMD from healthy controls. The highest sensitivity (0.864) was obtained by the model trained on the masked ILM-BM.

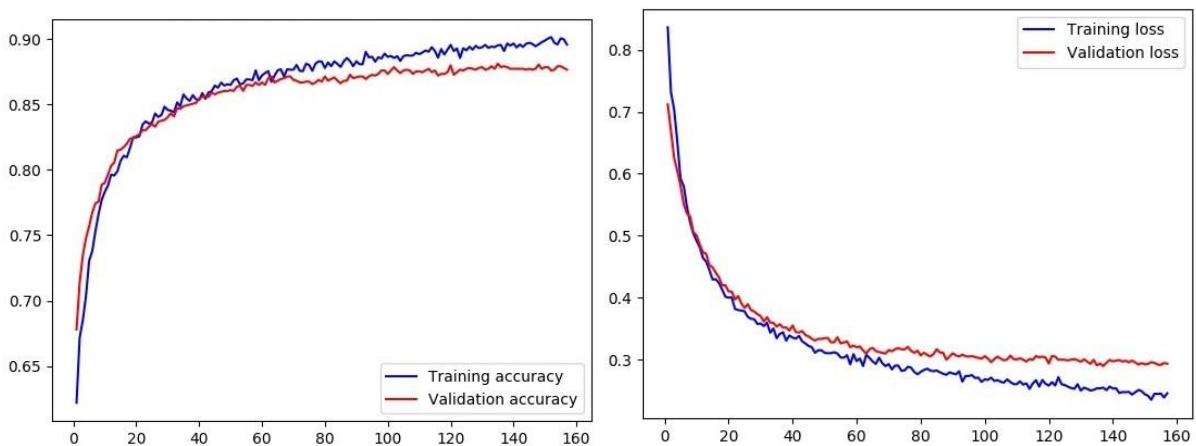

**Figure 7.** Accuracy (left) and loss (right) for the training (blue) and validation (red) on the whole image. The x-axis denotes the number of epochs.

When comparing the accuracy and AUROC for the masking and cropping methods, the masking ILM-BM had higher accuracy, sensitivity, and specificity than the cropping ILM-BM, whereas the latter had higher AUROC. For the rest of the cases, and except for the sensitivity of the model trained on BM-CHO, the results show that the models trained on masked images performed worse than the models using cropping of the equivalent area. The binary classifier trained on the RPE-BM mask yielded an AUROC of 0.932.

Table 2 shows the *p*-values resulting from the statistical analysis between the different models. There was not a statistically significant difference between the models trained on masked information of the RPE-BM and on the RPE-BM binary segmentation ($p$ = 0.18). The same was observed for the retina (ILM-BM) ROI between the masking and cropping approaches ($p$ = 0.47). Although there was a statistically significant difference between masking and cropping for the models trained on the ILM-BM and RPE-BM, or between the cropping RPE-BM and the RPE-BM mask, the *p*-values were not lower than 0.01. Lastly, it is noteworthy that statistically significant differences were observed between the model that obtained the highest AUROC (model trained on the whole B-scan image) and all other models.

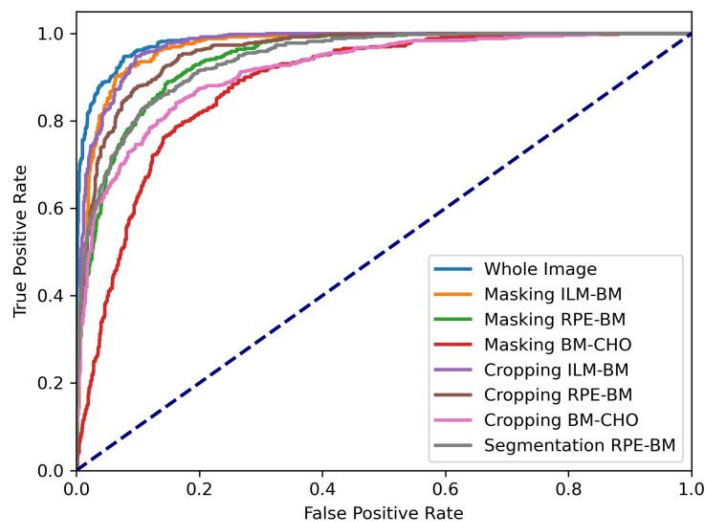

**Figure 8.** ROC curve for all models in the test set.

**Table 2.** Statistical comparison between the ROC curves for all models. The results correspond to the *p*-values obtained with the Delong test for unpaired data.

|  |  | Masking | | | Cropping | | | Segmentation-only |
|---|---|---|---|---|---|---|---|---|
|  |  | ILM-BM | RPE-BM | BM-CHO | ILM-BM | RPE-BM | BM-CHO | RPE-BM |
| Whole Image | IMG | 0.004 | <0.001 | <0.001 | 0.02 | <0.001 | <0.001 | <0.001 |
| Masking | ILM-BM |  | <0.001 | <0.001 | 0.47 | 0.03 | <0.001 | <0.001 |
|  | RPE-BM |  |  | <0.001 | <0.001 | 0.03 | 0.003 | 0.18 |
|  | BM-CHO |  |  |  | <0.001 | <0.001 | 0.006 | <0.001 |
| Cropping | ILM-BM |  |  |  |  | 0.004 | <0.001 | <0.001 |
|  | RPE-BM |  |  |  |  |  | <0.001 | <0.001 |
|  | BM-CHO |  |  |  |  |  |  | 0.09 |

## 4 Discussion

The model trained on the whole B-scan image has shown to be the best to differentiate between healthy controls and intermediate AMD subjects (AUROC = 0.983). Such results indicate that, in this specific classification problem, training on more information surpasses the performance of training on pre-selected ROIs, namely the retina, RPE-BM complex, and the choroid.

The models whose performance followed the best one were the cropping ILM-BM (AUROC = 0.974) and the masking ILM-BM (AUROC = 0.97). For both, a statistically significant difference was observed when compared to the best model ($p < 0.05$), but not between models ($p = 0.47$). Given the main difference between these two models and the best one (whole B-scan) is the choroidal information, we can conclude that the choroid retrieved from the OCT has an added value for the intermediate AMD classification with deep learning techniques. Such results support the observations reported by Srivastava et al. (Srivastava et al. 2020).

Due to the bounding box cropping, additional information outside of the RPE-BM may have been used in the training of the cropping RPE-BM model. This means that the masked images for the RPE-BM complex may not correspond to the same area as those images obtained with the cropping method. Also, the fact that the masking approach turns to zero all information outside the ROI, contributed to poorer performance of the masking approach since the percentage of information in the resulting image was considerably lower. Moreover, there is also a discrepancy in the scaling in the axial direction, since the cropping model rescales the information to fit the complete input, while the masking models keep the original proportions (e.g., less width for the RPE-BM region than for the ILM-BM). Such discrepancies may justify the difference observed between cropping and masking methods for the RPE-BM and BM-CHO ROIs.

When comparing the metrics for the masking and cropping methods, the conclusions were not straightforward, since the masking ILM-BM had higher accuracy, sensitivity, and specificity than the cropping ILM-BM, whereas the latter had higher AUROC. However, in this dataset, there were approximately two times more images belonging to AMD subjects than to healthy controls. Therefore, for the model evaluation, the AUROC is a better metric since it is appropriate for classification problems with class imbalance. The AUROC results show that the models trained and tested on masked images performed worse than the models using cropping of the equivalent area.

Although the AUROCs of the models trained on different ROIs were significantly lower than the one obtained for the whole OCT B-scan image, it is noteworthy that the obtained metrics were still high. All AUROCs were higher than 90% except for the masking BM-CHO (88.4%). Such results show that partial information of a standard retinal OCT B-scan can still be used for the intermediate stage AMD classification with an AUROC higher than 90%. This is of importance for not discarding OCT B-scans where only partial informative data is available due to imaging acquisition artifacts (e.g., partial cropping, eye movements, etc). Moreover, the performance of the model trained only on the BM-CHO is particularly interesting, as this part of the image does not have appreciable visual changes in the presence of intermediate AMD and, hence, it is not considered relevant in current clinical practice. The obtained results indicate that there is different information on the imaged choroid when comparing healthy controls to intermediate AMD subjects. Although it is not clear whether there is any clinically relevant information (i.e. pathologically related information), or whether the variability in this region occurs only because of imaging artifacts from the anterior layers, it is certain that the light transmitted to the choroid depends on the attenuation that happens in the retina. Therefore, any changes that may occur in the retina, such as the occurrence of drusen deposits, will certainly change how the light reaches the choroid, resulting in changes in the pixel intensity distribution. Therefore, any conclusions on the clinical relevance of the imaged choroid for the intermediate AMD classification should be tackled carefully. Still, this issue is more prominent the more advanced the disease is, when RPE atrophy appears, while most early to intermediate AMD cases should have minimal transillumination.

The AUROC score of the masked RPE-BM shows that the distinctive shape of the RPE is an important feature for the intermediate AMD classification. In fact, no statistically significant difference was observed when comparing the masking RPE-BM with the RPE-BM mask ($p$ = 0.18). Therefore, we may conclude that the pixel intensities within the RPE-BM complex do not have an added value in comparison with its morphology for the intermediate AMD classification using the proposed approach. However, it is worth noting that changes in reflectivity within this layer are still clinically relevant for underlying CNV and should not be overlooked.

Although the developed binary classification models yielded good AUROC scores, there are a few aspects that can be explored to further extend the current research. All presented models had the input images resized, as the VGG16 CNN requires the input to be of a specific size. This is especially relevant to the models trained on an ROI, as the real thickness of the layers is lost due to this resizing. One option that could be explored in the future is to add the average thickness of the layer as additional input to the network or adjust the model to be able to have the original images as input.

One limitation of this study is that an accurate segmentation of the choroid was not available, so an estimation based on the BM and normative choroidal thickness values was used. Using a more accurate mask for the BM-CHO model could improve the results, since it will ensure all the choroidal information is included. However, to accurately segment the choroid-sclera interface is still a challenging task due the light attenuation in the retina.
Another important aspect is that the dataset was labelled on patient level. However, some of the images belonging to intermediate AMD OCT volumes showed no signs of drusen. These B-scans appear to be visually "healthy", and it would be interesting, in future research, to observe how many of these cases correspond to a misclassification.

Lastly, the results of this study may have been affected by the decision of optimizing the hyperparameters used for all models in the IMG dataset. For this dataset, it was verified that the training was correct and completed without under- and over-fitting issues. To compare the effect of different ROIs, the model and hyperparameters had to be the same in all the experiments. Although the selected hyperparameters are quite generic, such a decision may have induced some bias in favour of the IMG model, and hence contributed to its superior performance compared to other models. This means that the AUROC for all models, besides the IMG model, are a "pessimistic" or "lower threshold" value.

## 5 Conclusion

In this work, we show that there are much more informative data in a standard OCT B-scan image than those visually inspected in clinical practice. Based on local features analysis, a DL method such as the one presented in this study can efficiently differentiate between intermediate AMD subjects and healthy controls. Although the whole OCT B-scan, including the retina and the choroid, leads to a better classification of intermediate AMD, it is possible to obtain high levels of performance using only partial information of the B-scan, such as the RPE-BM complex or the choroid. Lastly, we have observed that consistently with clinical practice, RPE-BM thickness/shape plays a major role in intermediate AMD classification with DL techniques.